\documentclass{aa}

\usepackage{txfonts}

\usepackage{graphicx}
\usepackage{array}
\usepackage{epsfig}

\usepackage{natbib}
\bibpunct{(}{)}{;}{a}{}{,}

\newcommand{\teff}{$T_\mathrm{eff}$}
\newcommand{\logg}{$\log g$}

\newcommand{\prot}{$P_\mathrm{rot}$}

\newcommand{\tz}{\raisebox{-0.4ex}{\scriptsize \,0}}
\newcommand{\tpn}{\raisebox{-0.4ex}{\scriptsize $\pm$1}}

\newcommand{\tp}{\raisebox{-0.4ex}{\scriptsize +1}}
\newcommand{\tn}{\raisebox{-0.4ex}{\scriptsize $-$1}}
\newcommand{\tpp}{\raisebox{-0.4ex}{\scriptsize +2}}
\newcommand{\tnn}{\raisebox{-0.4ex}{\scriptsize $-$2}}

\newcommand{\rwd}{$R_\mathrm{WD}$}

\newcommand{\bdip}{$B^\mathrm{d}_\mathrm{pol}$} 
\newcommand{\bqua}{$B^\mathrm{q}_\mathrm{pol}$}
\newcommand{\boct}{$B^\mathrm{o}_\mathrm{pol}$}
\newcommand{\thed}{$\Theta^\mathrm{d}$}
\newcommand{\phid}{$\Phi^\mathrm{d}$}
\newcommand{\theq}{$\Theta^\mathrm{q}$}
\newcommand{\phiq}{$\Phi^\mathrm{q}$}
\newcommand{\theo}{$\Theta^\mathrm{o}$}
\newcommand{\phio}{$\Phi^\mathrm{o}$}

\newcommand{\xoff}{${x'}_\mathrm{off}$}
\newcommand{\yoff}{${y'}_\mathrm{off}$}
\newcommand{\zoff}{${z'}_\mathrm{off}$}

\newcommand{\chisq}{$\chi^2$}
\newcommand{\chisqred}{$\chi^2_\mathrm{red}$}

\newcommand{\bpd}{\mbox{$B$--$\psi$ diagram}}
\newcommand{\bpds}{\mbox{$B$--$\psi$ diagrams}}

\newcommand{\lbc}{\mbox{$\lambda$--$B$ curves}}
\newcommand{\lbcsing}{\mbox{$\lambda$--$B$ curve}}

\newcommand{\halpha}{\mbox{H$\alpha$}}
\newcommand{\hbeta}{\mbox{H$\beta$}}
\newcommand{\hgamma}{\mbox{H$\gamma$}}

\newcommand{\sigmi}{$\sigma^-$}
\newcommand{\sigpl}{$\sigma^+$}

\newcommand{\glm}{$g_l^m$}
\newcommand{\hlm}{$h_l^m$}

\newcommand{\ohe}{\mbox{\object{HE\,1045$-$0908}}}
\newcommand{\opg}{\mbox{\object{PG\,1015+014}}}
\newcommand{\ogrw}{\mbox{\object{Grw\,$+70^{\circ}8247$}}}

\newcommand{\sdss}{\textit{Sloan Digital Sky Survey}}

\begin{document}

\title{Zeeman tomography of magnetic white dwarfs}
\subtitle{III.\ The \mbox{70--80} Megagauss magnetic field of \opg\thanks{
Based on observations collected at the European Southern Observatory,
Paranal, Chile, under programme ID \mbox{63.P-0003(A)}.}}

   \author{F.~Euchner\inst{1} \and
           S.~Jordan\inst{2} \and
           K.~Beuermann\inst{1} \and
           K.~Reinsch\inst{1} \and
           B.\,T.~G\"ansicke\inst{3}}

        \offprints{F.~Euchner, \email{feuchner@astro.physik.uni-goettingen.de}}
   
        \institute{Institut f\"ur Astrophysik, Universit\"at G\"ottingen,
        \mbox{Friedrich-Hund-Platz~1}, \mbox{37077~G\"ottingen}, Germany 
	\and 
	Astronomisches Rechen-Institut am ZAH, \mbox{M\"onchhofstr.~12--14}, 
	\mbox{69120~Heidelberg}, Germany 
	\and 
	Department of Physics, University of Warwick, Coventry~CV4~7AL, UK}
             
\date{Received 12~January 2006 / Accepted 31~January 2006}

\abstract
{} 
{ 
We analyse the magnetic field geometry of the magnetic DA white dwarf
\opg\ with our Zeeman tomography method.
}
{ 
This study is based on rotation-phase resolved optical flux and
circular polarization spectra of \opg\ obtained with FORS1 at the ESO VLT.
Our tomographic code makes use of an extensive database of pre-computed 
Zeeman spectra. The general approach has been described in Papers~I and II
of this series.
}
{ 
The surface field strength distributions for all rotational phases of
\opg\ are characterised by a strong peak at \mbox{70\,MG}.  A
separate peak at \mbox{80\,MG} is seen for
about one third
of the rotation cycle.
Significant contributions to the Zeeman features 
arise from regions with field strengths between 50 and \mbox{90\,MG}. 
We obtain equally good simultaneous fits to the observations,
collected in five phase bins, for two different field
parametrizations: (i) a superposition of individually tilted and
off-centred zonal multipole components; and (ii) a truncated multipole
expansion up to degree \mbox{$l = 4$} including all zonal and tesseral
components.
The magnetic fields generated by both parametrizations
exhibit a similar global structure of the absolute surface field values,
but differ considerably in the 
topology of the
field lines.
An effective photospheric
temperature 
of \mbox{\teff\ = 10\,000\,$\pm$\,1000\,K} was found.
}
{ 
Remaining discrepancies between the observations and our best-fit models 
suggest that additional small-scale structure of the magnetic field exists which 
our field models are unable to cover due to the restricted number of
free parameters.
}
\keywords{white dwarfs -- stars:magnetic fields -- stars:atmospheres
-- stars:individual (\opg) -- polarization}
   
\titlerunning{Zeeman tomography of magnetic white dwarfs.\ III.}
\authorrunning{F.~Euchner et al.}
\maketitle


\section{Introduction}

In about 170 of the 5448 white dwarfs (WDs) listed in the Web
Version\footnote{http://www.astronomy.villanova.edu/WDCatalog/index.html,
January~2006.} of the Villanova White Dwarf Catalog magnetic fields
between \mbox{2\,kG--1000\,MG} have been detected, corresponding to a
fraction of \mbox{$\simeq$\,3\,\%}
\citep{mccook+sion99-1,wickramasinghe+ferrario00-1,vanlandinghametal05-1}.
Most of the WDs have not been scrutinisingly examined for the presence
of a magnetic field, however, and a statistical study suggests that
the true fractional incidence could be as high as \mbox{20\,\%}\
\citep{liebertetal03-1,schmidtetal03-1}.
The magnetic white dwarfs (MWDs) are widely believed to be the successors of
the chemically peculiar magnetic Ap stars, which are the only main
sequence stars to show substantial globally organised magnetic fields.
However, this scenario is challenged by the recent detections of kilogauss-size
fields in several MWDs as well as in their direct progeny (central stars
of planetary nebulae and hot subdwarfs,
\citeauthor{aznarcuadradoetal04-1},
\citeyear{aznarcuadradoetal04-1},
\citeauthor{jordanetal05-1},
\citeyear{jordanetal05-1},
\citeauthor{otooleetal05-2},
\citeyear{otooleetal05-2}).
Undoubtedly, further theoretical and observational efforts are
required in order to shed more light on the role magnetic fields play
in the key stages of post-main sequence evolution.
For the present purpose, we consider MWDs as
stars displaying a field strength \mbox{$B \ga 1$\,MG}.

Due to the intrinsic faintness of WDs, \mbox{8-m} class telescopes are
required in order to record high-quality spectropolarimetric data
with sufficient time resolution as a basis for
studies of the magnetic field geometry.
In the course of our Zeeman tomography programme we have conducted
observations for a number of isolated (non-accreting) and accreting
MWDs at the ESO VLT with FORS1 in the spectropolarimetric mode.
%

\begin{figure*}[t]
\includegraphics[bb=16 40 564 742,width=8.8cm,clip]{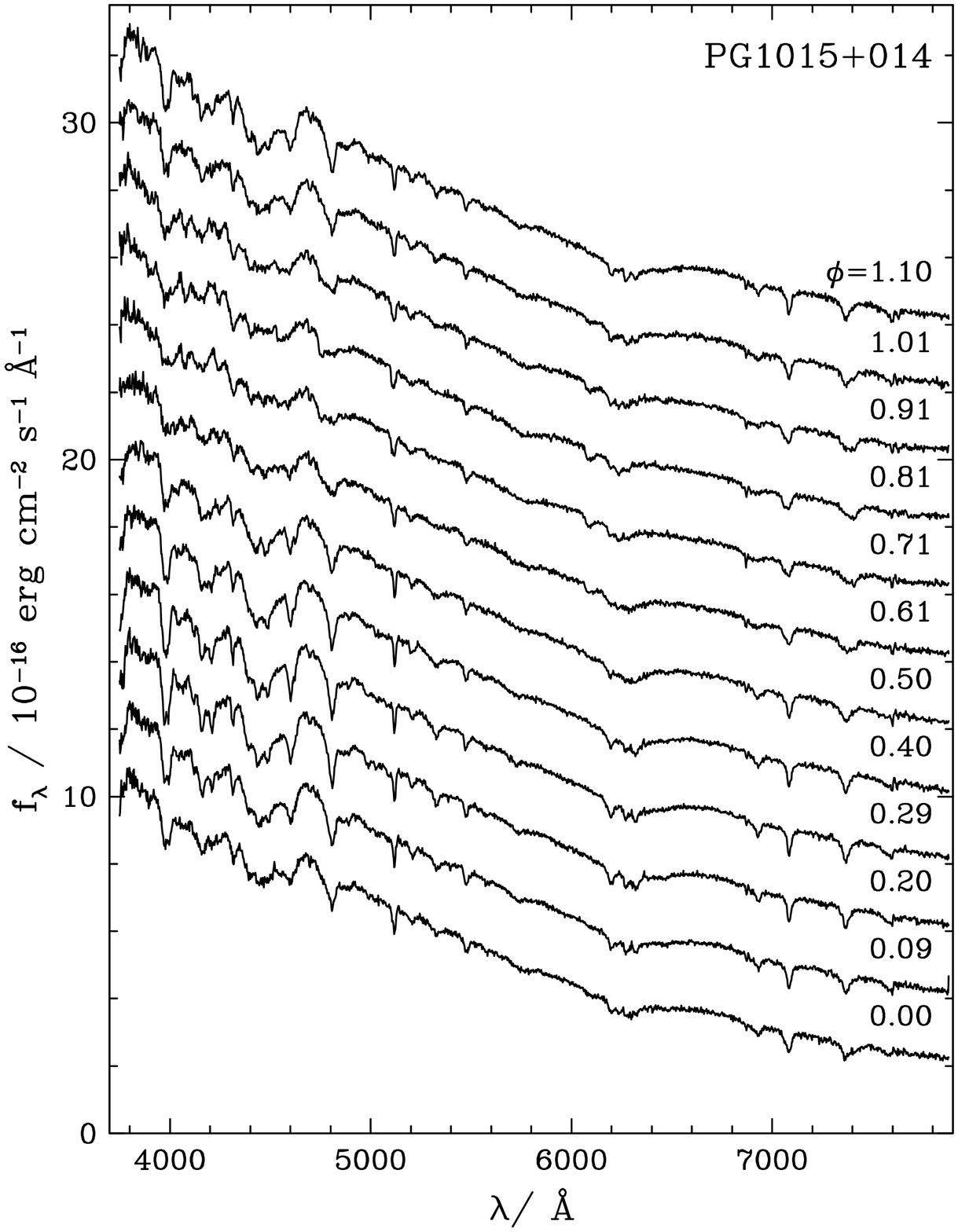}
\hfill
\includegraphics[bb=16 40 564 742,width=8.8cm,clip]{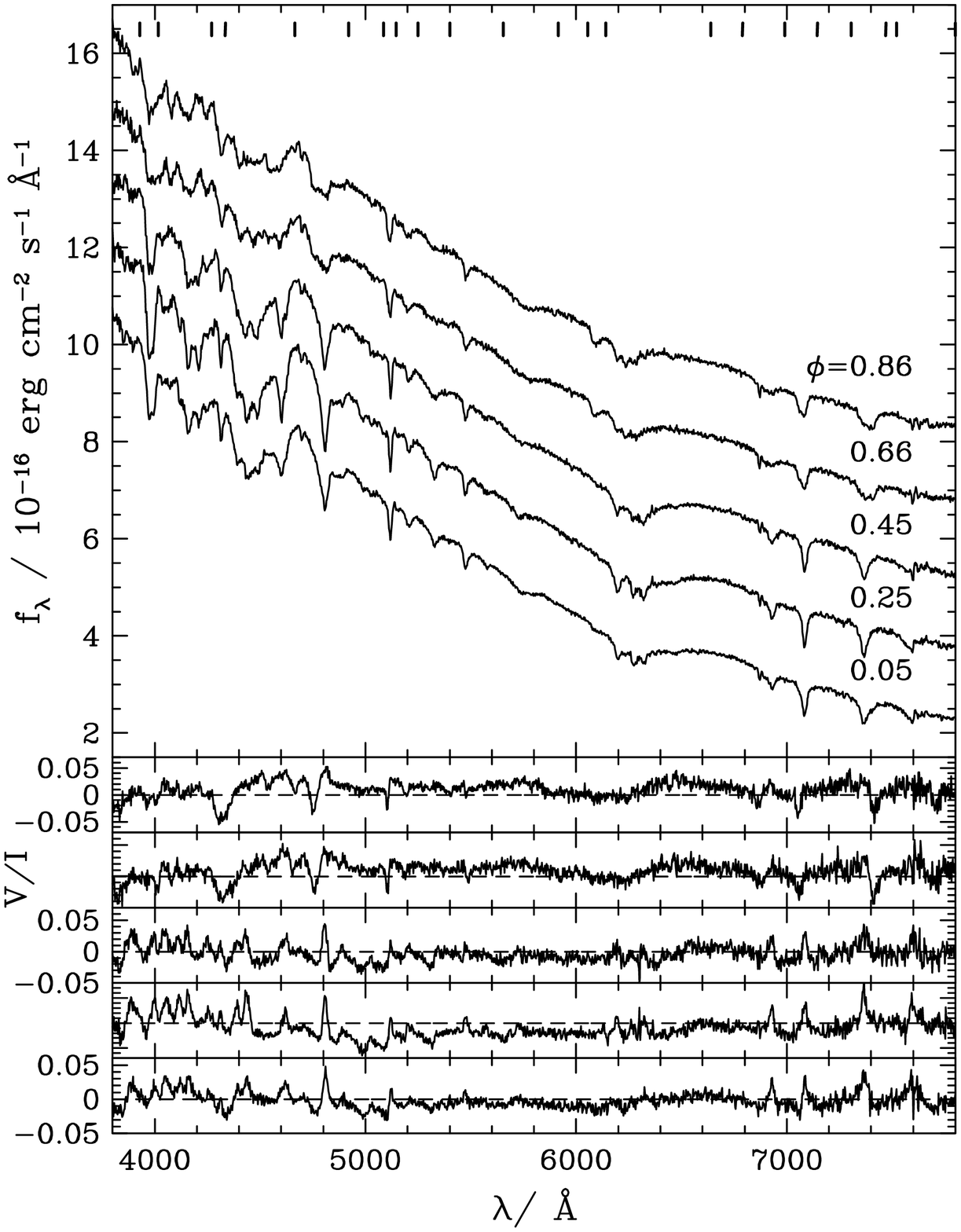}
\caption{\textit{Left panel:} Flux spectra of \opg\ from May~1999. The
uppermost 11 spectra have been shifted upwards by 2 flux units,
respectively.  \textit{Right panel:} Combined flux and circular polarization 
spectra of \opg\ from
the May~1999 observations. These spectra, which have been collected into five
phase bins, will be used as input spectra for the Zeeman tomography
procedure. For clarity, the uppermost four flux spectra have been shifted
upwards by 1.5 flux units, respectively. The tick marks below the top
axis indicate the wavelengths that were used to adjust the continuum flux level
of the the model spectra to the observations.}
\label{fig:pg1015-obs-full}
\end{figure*}

In the present work -- the third paper of our series on Zeeman
tomography  -- we present an application of our code to
observations of the non-accreting white dwarf \opg\ and find a field
geometry that deviates strongly from centred dipoles or quadrupoles.
In the first paper of the series, we
have demonstrated that Zeeman tomography is a suitable method to 
recover field geometries by analysing synthetically generated 
spectra \citep[][ hereafter Paper~I]{euchneretal02-1}.
In a first application of this theory,
we have derived a quadrupole-dominated field
structure with a prevailing field of \mbox{$\simeq$\,16\,MG} for 
\ohe\ \citep[][ hereafter Paper~II]{euchneretal05-2}.

The magnetic DA white dwarf \opg, 
discovered in the Palomar Green survey \citep{greenetal87-1},
was observed by \citet[][ hereafter WC88]{wickramasinghe+cropper88-1}
with the RGO spectrograph at the AAT 
in the wavelength range \mbox{4000--7000\,\AA}.
Their phase-resolved spectroscopy and low-resolution circular polarimetry
revealed significant modulations in flux and circular polarization ($V/I$)
over the rotation cycle.
From nearly sinusoidal oscillations of the wavelength-averaged degree of 
circular polarization 
between \mbox{$-$1.5\,\%} and \mbox{1.5\,\%} the authors derived 
a rotational period of
\mbox{\prot\ = 98.7\,min}, which was later confirmed with 
higher accuracy by \citet{schmidt+norsworthy91-1}
who used white-light circular polarimetry.
In the individual polarization spectra of WC88, 
$|V/I|$ is \mbox{$\simeq$\,5\,\%} in the
continuum and up to \mbox{$\simeq$\,10\,\%} in individual features.
They fitted theoretical MWD model 
spectra to the observations and 
found an obliquely 
rotating magnetic dipole model with a polar field strength of 
\mbox{\bdip\ = 120\,$\pm$\,10\,MG} and an almost 
equator-on view to be the best-fitting field geometry. 
Remaining discrepancies between observations and model 
spectra were attributed
to higher-field regions superimposed on the dipolar field structure.
Our analysis provides a substantially improved insight into the
field structure of \opg.


\section{Observations}
\label{sec:observations}

We have obtained spin-phase resolved circular spectropolarimetry of
\opg\ with FORS1 
at the ESO VLT on May 15, 1999. 
The spectrograph was operated in spectropolarimetric (PMOS) mode,
with the GRIS\_300V+10 grism and an order separation filter
GG~375, yielding a usable wavelength range \mbox{$\sim$\,3850--7900\,\AA}.
With a slit width of 1\arcsec\, the FWHM spectral
resolution was \mbox{13\,\AA}\ at \mbox{5500\,\AA}.
The observational data have been reduced according to standard
procedures (bias, flat field, night sky subtraction, wavelength
calibration, atmospheric extinction, flux calibration) using the
context MOS of the ESO MIDAS package.
The instrumental setup and the data reduction are analogous 
to those employed for our analysis of \ohe\ (Paper~II).

\begin{figure*}[t]
\includegraphics[bb=18 40 564 742,width=8.8cm,clip]{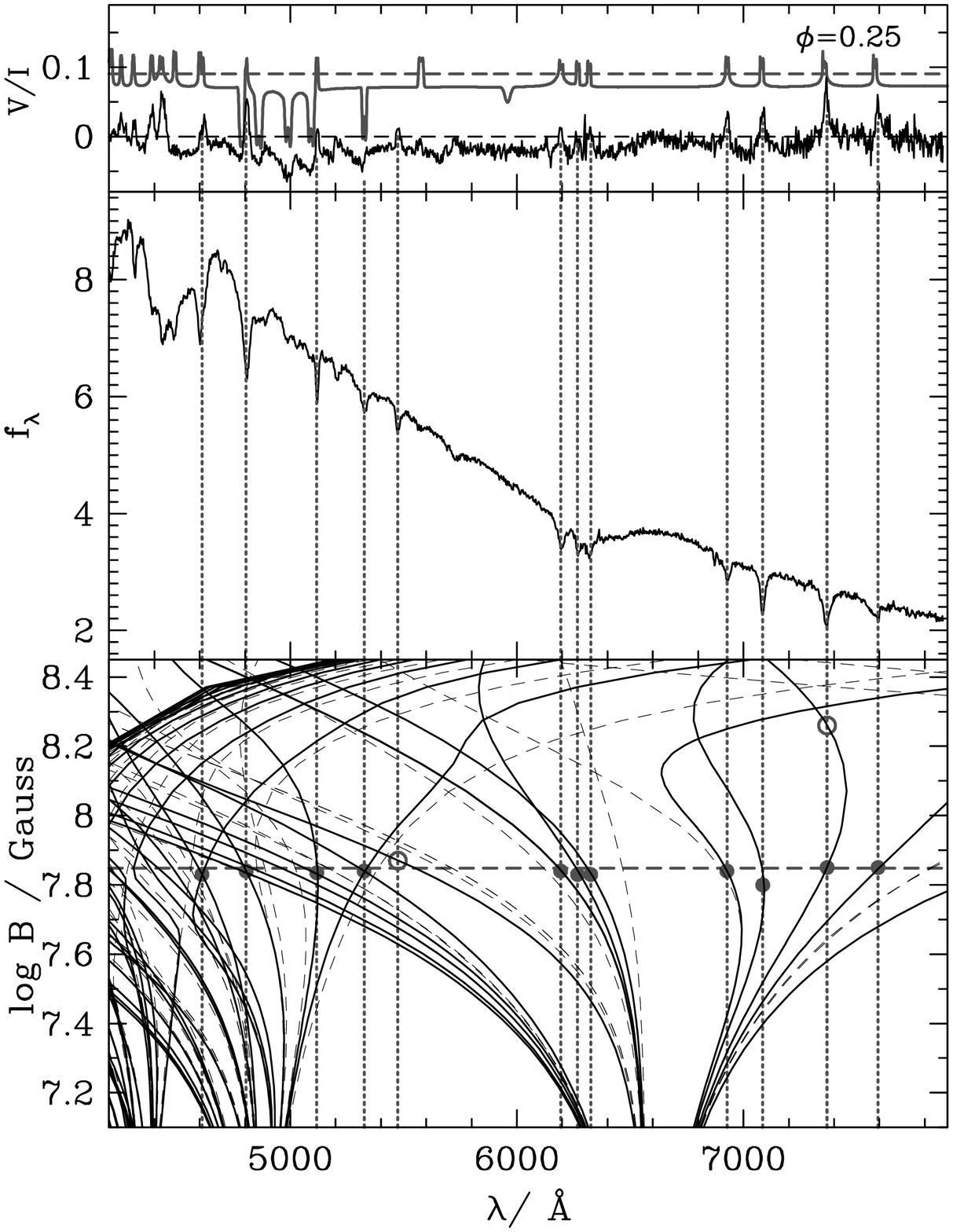}
\hfill
\includegraphics[bb=18 40 564 742,width=8.8cm,clip]{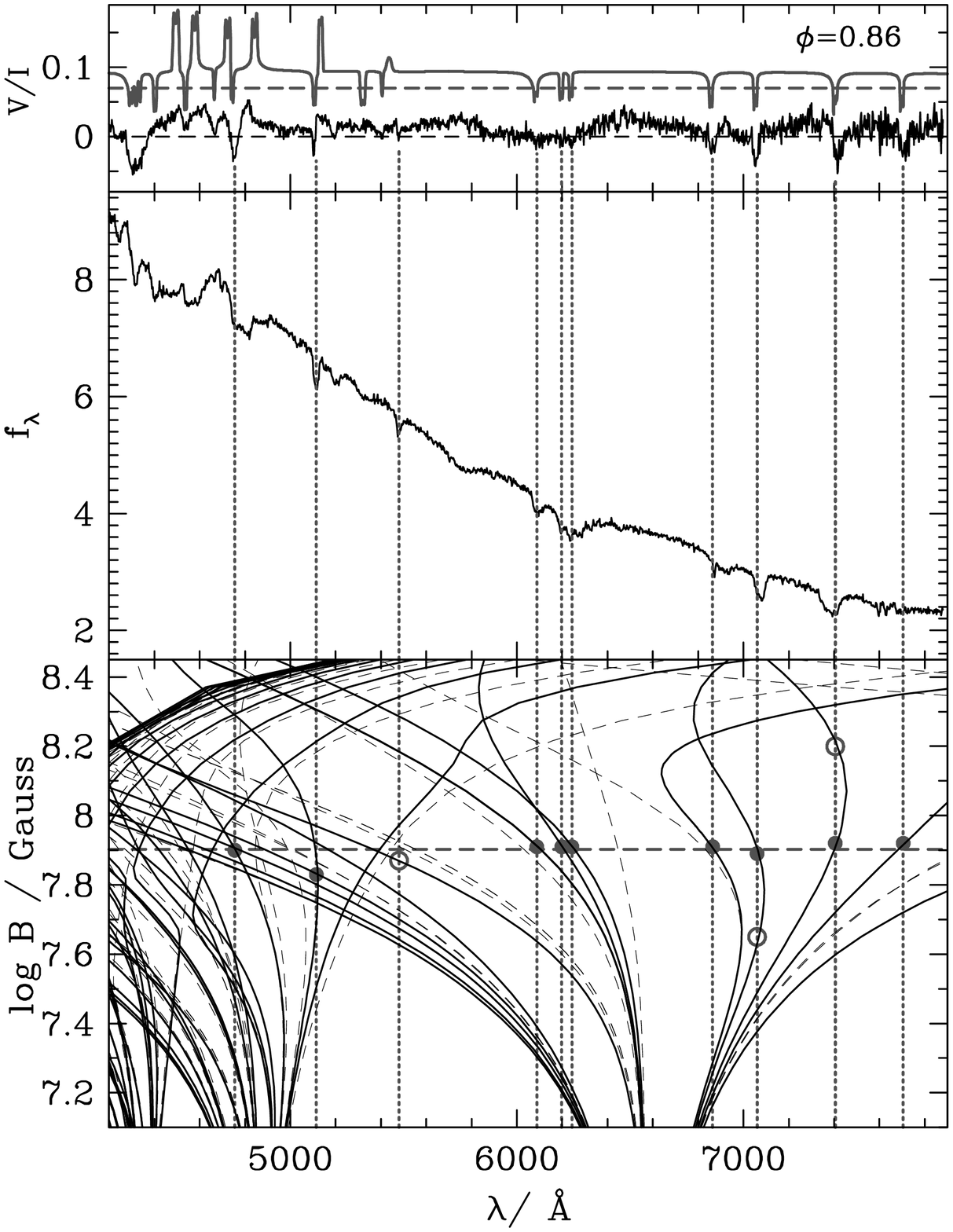}
\caption{Zeeman maximum (\mbox{$\phi = 0.25$}, \textit{left panel})
and minimum (\mbox{$\phi = 0.86$}, \textit{right panel}) flux
($f_\lambda$) and circular polarization ($V/I$) spectra for \opg,
plotted along with the theoretically predicted field-dependent
transition wavelengths for Balmer absorption lines (\lbc).
Transitions corresponding formally to \mbox{$|\Delta l| = 0, 2, 4,
\dots$} with the zero-field angular momentum quantum number $l$ have
been plotted as dashed curves in the bottom parts of the figures.
Filled circles denote unambiguous identifications of transitions, open
circles less certain ones.
In the top parts, theoretical circular polarization spectra are shown
for comparison (shifted upwards for clarity, see the text for the
adopted model parameters).
$f_\lambda$ is given in units of \mbox{$10^{-16}$ erg cm$^{-2}$
s$^{-1}$ $\AA^{-1}$}.}
\label{fig:obs-zeeman-spaghetti}
\end{figure*}

We secured a sequence of 12 exposures with an exposure time of 8~min each, 
covering a full rotation cycle. 
We were able to reach a signal-to-noise ratio \mbox{$S/N \simeq 100$} for the 
individual flux spectra (Fig.~\ref{fig:pg1015-obs-full}, \textit{left panel}).
The wavelength-dependent degree of circular polarization ($V/I$) was
computed from two consecutive exposures -- 
with the retarder plate positions differing by \mbox{90\degr}\ --
in order to eliminate Stokes
parameter crosstalk, thus yielding six independent phase bins.
The flux spectra do not show the typical absorption signature of low-
and intermediate-field MWDs with the \mbox{\halpha\ $\pi$} component
at the zero-field wavelength surrounded by broader \sigmi\ and \sigpl\
troughs, but exhibit a variety of distinct sharp lines scattered over
the whole optical range which vary in strength, position, and shape
with the rotational phase.
The continuum circular polarization differs significantly from zero for
most phases. 
Both phenomena are characteristic of a high-field object with 
\mbox{$B \ga 50$\,MG} \citep[e.g.][]{wickramasinghe+ferrario00-1}.

We collect the observed flux and circular polarization spectra into
five nearly-equidistant phase bins (\mbox{$\phi = 0.05$}, 0.25, 0.45,
0.66, 0.86).  The phases refer to the ephemeris of
\citet{schmidt+norsworthy91-1} with \mbox{$\phi = 0$} denoting the
change from positive to negative continuum polarization\footnote{These
authors used a different convention for the sign of the circular
polarization than we did, since in their data $V/I$ changes from negative
to positive values at \mbox{$\phi = 0$}.}.
These flux and polarization spectra (Fig.~\ref{fig:pg1015-obs-full},
\textit{right panel}) form the basis of our tomographic analysis.
We investigate two particular phases in greater detail: 
\mbox{$\phi = 0.25$}, when the Zeeman features are strongest 
(``Zeeman maximum''); and
\mbox{$\phi = 0.86$}, when they are weakest (``Zeeman minimum'').


\section{Qualitative analysis of the magnetic field geometry}
\label{sec:qualitative}

As a first step of the analysis, we compare the positions and
strengths of the most prominent features in the observed spectra with
the expected field-dependent wavelengths of the hydrogen transitions
$\lambda^{\textrm{H}}(B)$, henceforth referred to as \lbc\
\citep{forsteretal84-1, roesneretal84-1, wunneretal85-1}.
In Fig.~\ref{fig:obs-zeeman-spaghetti}, we show the observed flux and
circular polarization spectra along with the \lbc\ for \mbox{$\phi =
0.25$} \textit{(left panel)} and 0.86 \textit{(right panel)}.
Transitions that could be unambiguously identified with specific
spectral features have been marked with filled grey circles and are
listed in Table~\ref{tab:zeeman-transitions}. Possible or unlikely
identifications are displayed as open circles.
In the presence of an electric field additional transitions can occur
which formally correspond to \mbox{$|\Delta l| = 0, 2, 4,
\dots$}, with $l$ being the angular momentum quantum number in the
zero-field nomenclature. These components have been plotted as 
dashed curves in the bottom parts of
Fig.~\ref{fig:obs-zeeman-spaghetti}.

\begin{table}[t]
\caption{Positions of observed Zeeman features and corresponding Balmer 
transitions. A colon denotes a less certain identification.
}
\label{tab:zeeman-transitions}
\centering
\begin{tabular}{llll} 

\hline \hline 
\noalign{\smallskip}

$\lambda$ / \AA\ & Transition & $\lambda$ / \AA\ & Transition \\ 

\noalign{\smallskip} 
\hline
\noalign{\smallskip}

\multicolumn{2}{l}{\textit{(\mbox{$\phi = 0.25$})}} & 
\multicolumn{2}{l}{\textit{(\mbox{$\phi = 0.86$})}}\\

4610 & \hbeta\,(2s\tz\,$\rightarrow$\,4f\tz)    & 4755 & \hbeta\,(2p\tz\,$\rightarrow$\,4d\tn)  \\
     & \hbeta\,(2p\tp\,$\rightarrow$\,4s\tz)    & 5115 & \hbeta\,(2s\tz\,$\rightarrow$\,4f\tn)  \\
4805 & \hbeta\,(2p\tz\,$\rightarrow$\,4d\tn)    & 5480 & \halpha\,(2p\tz\,$\rightarrow$\,3s\tz)\,:  \\
5117 & \hbeta\,(2s\tz\,$\rightarrow$\,4f\tn)    & 6088 & \halpha\,(2p\tpn\,$\rightarrow$\,3d\tpn) \\
5326 & \halpha\,(2p\tn\,$\rightarrow$\,3d\tz)   & 6198 & \halpha\,(2s\tz\,$\rightarrow$\,3p\tz) \\ 
     & \hbeta\,(2p\tp\,$\rightarrow$\,4d\tz)    & 6243 & \halpha\,(2p\tz\,$\rightarrow$\,3d\tz)  \\
5475 & \halpha\,(2p\tz\,$\rightarrow$\,3s\tz)\,:   & 6864 & \halpha\,(2p\tp\,$\rightarrow$\,3s\tz) \\
6193 & \halpha\,(2p\tpn\,$\rightarrow$\,3d\tpn) & 7060 & \halpha\,(2s\tz\,$\rightarrow$\,3p\tn) \\
6268 & \halpha\,(2s\tz\,$\rightarrow$\,3p\tz)   & 7405 & \halpha\,(2p\tn\,$\rightarrow$\,3d\tnn) \\
6326 & \halpha\,(2p\tz\,$\rightarrow$\,3d\tz)   & 7705 & \halpha\,(2p\tz\,$\rightarrow$\,3d\tn) \\
6927 & \halpha\,(2p\tp\,$\rightarrow$\,3s\tz)   &  & \\
7085 & \halpha\,(2s\tz\,$\rightarrow$\,3p\tn)   &  & \\
7369 & \halpha\,(2p\tn\,$\rightarrow$\,3d\tnn)  &  & \\
7594 & \halpha\,(2p\tz\,$\rightarrow$\,3d\tn)   &  & \\
 
\noalign{\smallskip} 
\hline

\end{tabular}
\end{table}

In the Zeeman maximum spectrum \mbox{($\phi = 0.25$)}, three
\mbox{H$\alpha$ $\pi$} and four \mbox{H$\alpha$ $\sigma^+$}
transitions with \mbox{$\lambda > 6000$\,\AA}\ are clearly visible as
positive peaks superimposed on a negative polarization continuum. In
particular, the narrow feature at \mbox{7085\,\AA} which corresponds to a
local extremum of the \mbox{2s\tz\,$\rightarrow$\,3p\tn}\ \lbcsing\
allows a reliable estimation of the prevailing field strength.
For \mbox{$\lambda < 6000$\,\AA}, we identify three strong features
showing positive polarization (at 4610, 4805, and \mbox{5117\,\AA})
with \mbox{H$\beta$ $\sigma^+$} transitions.
Our identifications of several line features differ from those
suggested by WC88 for the same rotational phase.
The feature at \mbox{5326\,\AA}\ shows a noticeable negative
polarization and is interpreted as a blend of a \mbox{H$\alpha$
$\sigma^-$} and a \mbox{H$\beta$ $\sigma^+$} component, with the
\mbox{2p\tn\,$\rightarrow$\,3d\tz}\ transition dominating. The strong
line at \mbox{5475\,\AA}\ with positive polarization probably arises
from the \mbox{H$\alpha$ $\pi$} transition
\mbox{2p\tz\,$\rightarrow$\,3s\tz}.
The feature at \mbox{5930\,\AA}\ reported by WC88 is not present in our data.
Furthermore, there is a distinct feature with a positive peak in
polarization at \mbox{5200\,\AA}\ which they attribute to the
\mbox{H$\alpha$ $\sigma^-$} transition
\mbox{2p\tp\,$\rightarrow$\,3d\tpp}. We question this identification
because the \mbox{H$\alpha$ $\sigma^-$} components produce lines with
strong negative polarization.
However, no obvious match with a specific \lbcsing\ can be found in
Fig.~\ref{fig:obs-zeeman-spaghetti}, so the origin of 
this feature remains unclear.
We do not attempt to identify spectral features at wavelengths
\mbox{$\lambda < 4600$\,\AA} because the number of candidate
transitions from the overlapping \hbeta\ and \hgamma\ manifolds is too large.
Taking into account all these identifications, we conclude that the
distribution of the corresponding field strengths must be centred
quite sharply at \mbox{$\simeq$\,70\,MG}. To illustrate this, we show
in the top part of 
Fig.~\ref{fig:obs-zeeman-spaghetti} \textit{(left panel)} 
a theoretical circular polarization
spectrum for a \emph{single} value \mbox{$B = 69$\,MG} with
\mbox{$\psi = 51$\degr}, and \mbox{$T_\mathrm{eff} = 10$\,000\,K},
where $\psi$ denotes the viewing angle between
the magnetic field direction and the line of sight.

In the Zeeman minimum spectrum \mbox{($\phi = 0.86$)}, the absorption
signature for \mbox{$\lambda > 6000$\,\AA}\ is similar to that at
Zeeman maximum, although the \mbox{H$\alpha$ $\pi$} line components
are not visible as separate peaks in $V/I$, but rather produce a broad
depression of the overall positive continuum polarization.  The
\mbox{H$\alpha$ $\sigma^+$} components are wider than those at \mbox{$\phi =
0.25$}.  Both effects suggest a broader distribution of field
strengths at Zeeman minimum.
A theoretical circular polarization spectrum for a single value
\mbox{$B = 82$\,MG}, \mbox{$\psi = 129$\degr}, and
\mbox{$T_\mathrm{eff} = 10$\,000\,K} provides a good indication for
the prevailing field strength and direction.
As can be seen from the change of polarity in the continuum circular
polarization, the net magnetic field direction has changed sign along
the line of sight.
The fact that for both phases, which are separated by about half a
rotation cycle (\mbox{$\Delta \phi = 0.39$}), the field distributions
seem to be clearly peaked at values differing by only
\mbox{$\simeq$\,20\,\%} indicates a field geometry more complex than
centred or off-centred dipoles or quadrupoles.

\begin{figure}[t]
\includegraphics[bb=18 40 564 742,width=8.8cm,clip]{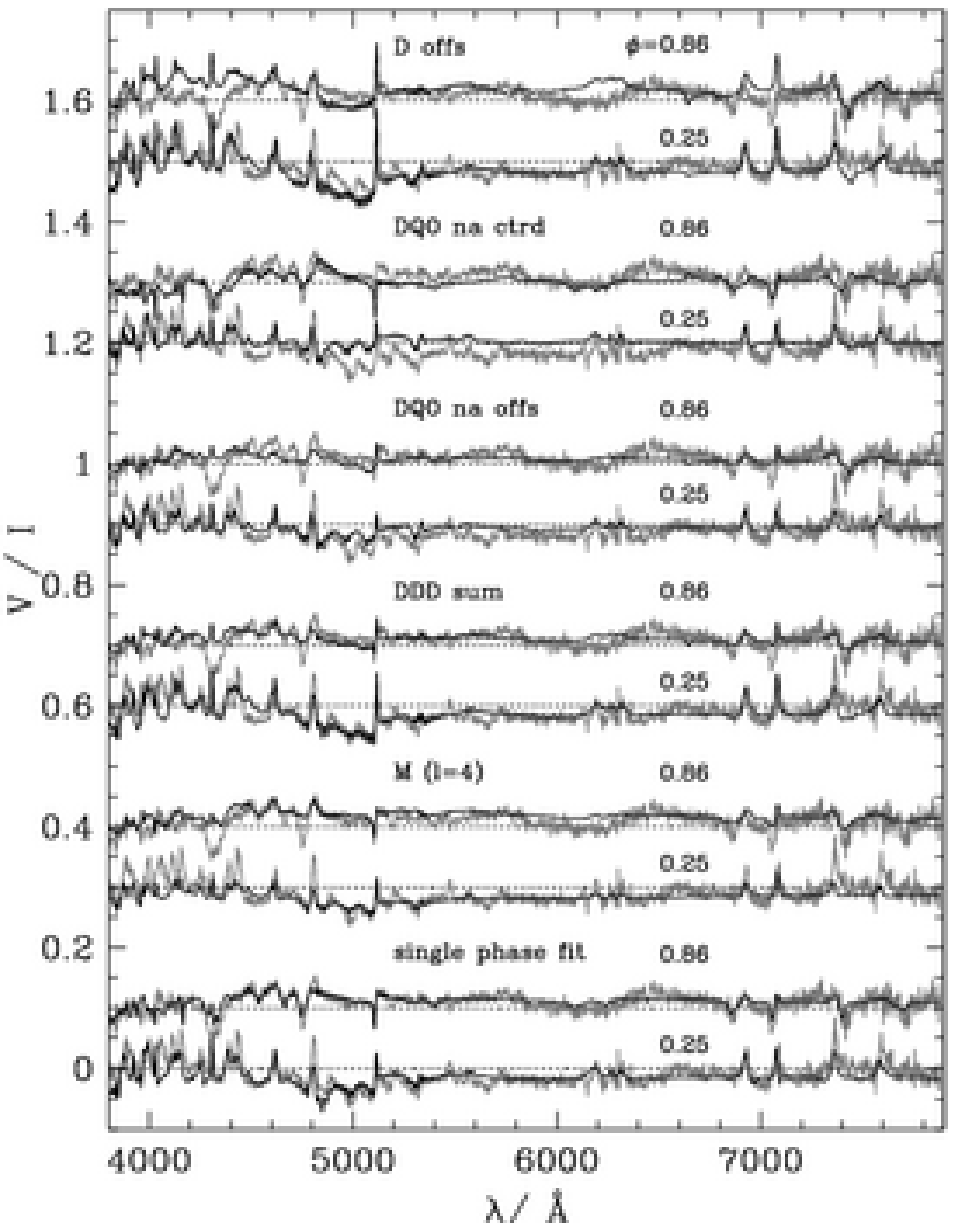} 
\caption{Observed circular polarization spectra for phases
\mbox{$\phi = 0.25$} and 0.86 (grey curves) and best-fit synthetic spectra
(black curves) for different parametrizations of the magnetic field
geometry. From top to bottom: off-centred dipole (D~offs);
centred, non-aligned combination of dipole,
quadrupole, and octupole (DQO~na~ctrd); 
off-centred, non-aligned
combination of dipole, quadrupole, and octupole (DQO~na~offs);
superposition of three individually off-centred, non-aligned dipoles (DDD~sum);
truncated multipole expansion up to degree \mbox{$l = 4$}.
The lowermost two curves show individual fits to single phases with a truncated
multipole expansion up to $l=5$ (\mbox{$\phi = 0.25$}) and $l=4$ 
(\mbox{$\phi = 0.86$}).
All curves except for the bottom one have been shifted vertically by
suitable amounts in $V/I$, with the horizontal dashed lines indicating
the respective levels of zero polarization.}
\label{fig:pg1015-multisp}
\end{figure}

\begin{table*}[t]
\caption{Best-fit magnetic parameters for the different
parametrizations of the magnetic field shown in
Fig.~\ref{fig:pg1015-multisp}. The uncertainties in the last digit
are denoted by the values in brackets. 
For each model, a short description is given in the first line. 
The best-fit inclinations are \mbox{24\degr\,$\pm$\,3\,\degr} (model~1), 
\mbox{45\degr\,$\pm$\,3\,\degr} (model~2), and 
\mbox{36\degr\,$\pm$\,4\,\degr} (model~3).
The superscripts `d', `q', and `o' refer to dipole, quadrupole, and
octupole, the subscript `pol' to the polar field strength. The last three
columns give the offsets in units of the white dwarf radius.
Formulae for the field structure are given in Paper~I.}
\label{tab:pg1015-fit-parameters}
\centering
\begin{tabular}{@{\hspace{5mm}}rrrrrrrrrrrr} 
\hline \hline \noalign{\smallskip}

\bdip & \thed & \phid & \bqua & \theq & \phiq & 
\boct & \theo & \phio & \xoff & \yoff & \zoff \\

(MG) & (\degr) & (\degr) & (MG) & (\degr) & (\degr) &
(MG) & (\degr) & (\degr) & (\rwd) & (\rwd) & (\rwd) \\ 

\noalign{\smallskip} \hline
\noalign{\smallskip}

\multicolumn{12}{l}{\textit{(1) D offs (off-centred dipole)}} \\

 $-$97\,(2) & 85\,(3) & 77\,(4) &  
 \multicolumn{1}{c}{--} & \multicolumn{1}{c}{--} & \multicolumn{1}{c}{--} &
 \multicolumn{1}{c}{--} & \multicolumn{1}{c}{--} & \multicolumn{1}{c}{--} & 
 $-$0.113\,(5) & $-$0.0036\,(2) &  0.162\,(4)  \\ [0.7ex]

 \multicolumn{12}{l}{\textit{(2) DQO na ctrd (non-aligned, centred combination of dipole, quadrupole, and octupole)}} \\

 $-$1.4\,(1) & 53\,(13) & 325\,(6) &  
 13.9\,(6) & 26\,(2) & 115\,(5) &
 174\,(1) & 65\,(3) & 94\,(5) & 
 \multicolumn{1}{c}{--} & \multicolumn{1}{c}{--} & \multicolumn{1}{c}{--} \\ [0.7ex]

 \multicolumn{12}{l}{\textit{(3) DQO na offs (non-aligned, off-centred combination of dipole, quadrupole, and octupole)}} \\

 $-$38\,(2) & 85\,(3) & 82\,(5) &  
 $-$15\,(3) & 85\,(5) & 21\,(3) &
 171\,(6) & 75\,(6) & 102\,(5) & 
 0.060\,(5) & 0.011\,(1) &  0.081\,(4) \\ [0.7ex]

\noalign{\smallskip} \hline
 
\end{tabular}
\end{table*}

\begin{table}[t]
\caption{Best-fit magnetic parameters for 
a superposition of three individually off-centred, non-aligned dipoles
(labelled DDD~sum in Fig.~\ref{fig:pg1015-multisp}).
The individual dipole components are denoted by D1, D2, and D3.
The uncertainties in the last digit
are given by the values in brackets.
The best-fit inclination is \mbox{23\degr\,$\pm$\,4\,\degr}.
}
\label{tab:pg1015-fit-parameters-ddd}
\centering
\begin{tabular}{@{\hspace{5mm}}crrr} 
\hline \hline \noalign{\smallskip}

  & D1 & D2 & D3 \\ 

\noalign{\smallskip} \hline
\noalign{\smallskip}

 \bdip\ (MG)    & $-$40\,(2) & 92\,(5)    & $-$38\,(3) \\ 
 \thed\ (\degr) & 44\,(4) & 63\,(2) & 63\,(5)    \\ 
 \phid\ (\degr) & 339\,(2) & 276\,(6) & 134\,(3) \\ 
 \xoff\ (\rwd)  & 0.04\,(1) & $-$0.012\,(2) & 0.27\,(1)     \\ 
 \yoff\ (\rwd)  & 0.35\,(3) & $-$0.136\,(8) & 0.080\,(7)   \\ 
 \zoff\ (\rwd)  & 0.33\,(1) & $-$0.28\,(3) & 0.21\,(2) \\  [0.7ex]
 
\noalign{\smallskip} \hline

\end{tabular}
\end{table}


\section{Zeeman tomography of the magnetic field}
\label{sec:tomography}

Our Zeeman tomography synthesises the observed spectra in a best-fit
approach. It makes use of an extensive pre-computed database of
theoretical flux and circular polarization spectra of magnetic white
dwarf atmospheres, with $B$, $\psi$, \teff, \logg, and the direction
cosine $\mu = \cos \vartheta$ as free parameters, where $\vartheta$
denotes the angle between the normal to the surface and the line of
sight.
The three-dimensional grid of \mbox{46\,800} Stokes~$I$ and $V$ profiles 
covers
400 $B$ values (1--400\,MG, in 1\,MG steps), nine $\psi$ values
(equidistant in \mbox{$\cos \psi$}), and 13 temperatures
\mbox{(8000--50\,000\,K)} for fixed \mbox{\logg\ = 8} and 
\mbox{$\mu = 1$}.
Limb darkening is accounted for in an approximate way by the linear
interpolation law ~\mbox{$I_{\lambda}(\mu) / I_{\lambda,\mu = 1} = a +
b \mu$}~ with constant coefficients \mbox{$a = 0.53$} and \mbox{$b =
0.47$}.
The best fit to the absolute flux distribution of \opg\ was obtained
with an effective temperature of \mbox{\teff\ = 10\,000 $\pm$ 1000\,K}.
We adopt this value in the subsequent analysis.
%

\begin{table}[t]
\caption{Best-fit magnetic parameters for 
a truncated multipole expansion up to degree \mbox{$l = 4$}
(labelled~M in Fig.~\ref{fig:pg1015-multisp}).
The coefficients \glm\ and \hlm\ are in MG.
The best-fit inclination is \mbox{47\degr}. The tilt and the azimuth
of the multipole axis relative to the rotational axis are
\mbox{22\degr}\ and \mbox{191\degr}.
}
\label{tab:pg1015-fit-parameters-multi}
\centering
\begin{tabular}{@{\hspace{5mm}}ccrrrr} 
\hline \hline \noalign{\smallskip}

 $m$ &  & $l=1$ & 2 & 3 & 4 \\ 
             
\noalign{\smallskip} \hline
\noalign{\smallskip}

 0 & \glm & 3.0 & 0.6 & 9.0 & 2.4 \\
 1 & \glm & $-$12.5 & 19.0 & $-$1.0 & 4.7 \\
   & \hlm & $-$28.2 & 7.1 & $-$15.6 & 11.0 \\
 2 & \glm & \multicolumn{1}{c}{--} & $-$19.6 & 11.6 & $-$10.2 \\
   & \hlm & \multicolumn{1}{c}{--} & $-$15.6 & $-$2.4 & 6.4 \\
 3 & \glm & \multicolumn{1}{c}{--} & \multicolumn{1}{c}{--} & 4.1 & $-$1.7 \\
   & \hlm & \multicolumn{1}{c}{--} & \multicolumn{1}{c}{--} & 14.9 & $-$6.9 \\
 4 & \glm & \multicolumn{1}{c}{--} & \multicolumn{1}{c}{--} & \multicolumn{1}{c}{--} & $-$4.7 \\
   & \hlm & \multicolumn{1}{c}{--} & \multicolumn{1}{c}{--} & \multicolumn{1}{c}{--} & $-$8.8 \\ [0.7ex]
 
\noalign{\smallskip} \hline

\end{tabular}
\end{table}

\begin{figure*}[t]
\includegraphics[bb=32 168 564 690,width=18cm,clip]{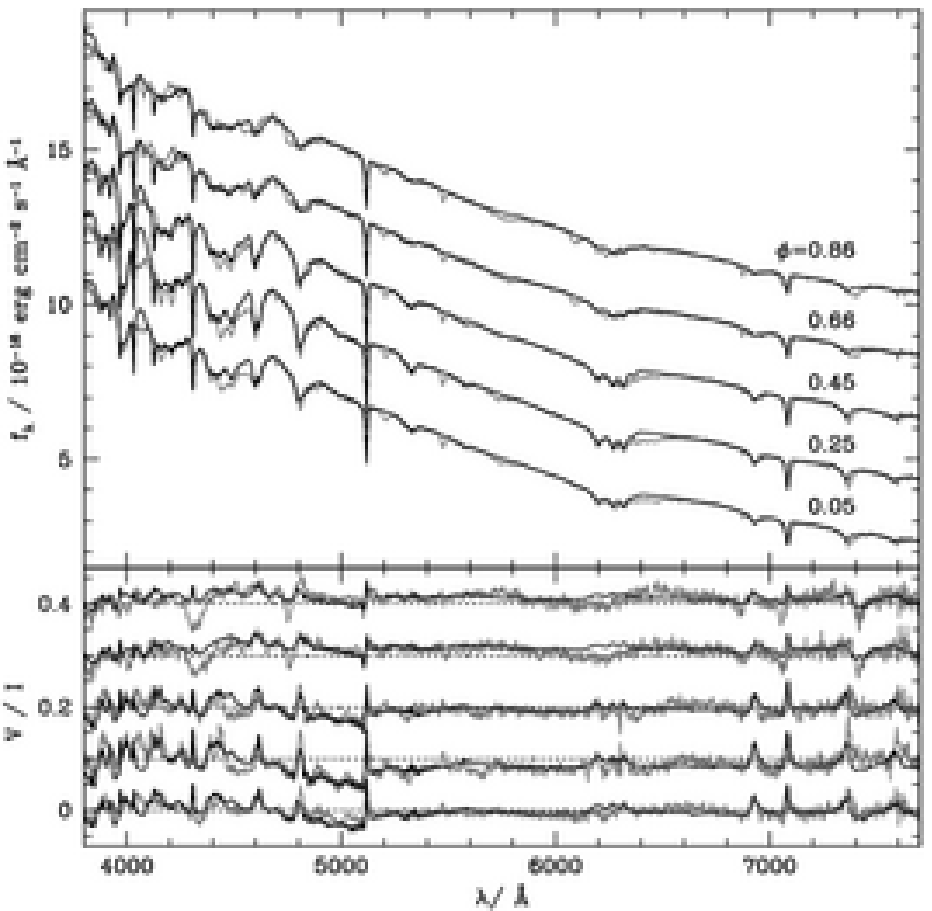}
\caption{Observed spectra of \opg\ (grey curves)
and best-fit synthetic spectra (black curves)
using a superposition of three individually off-centred, non-aligned
dipoles.
The upper four flux
(circular polarization) spectra have been shifted for clarity by 2, 4, 6, 
and 8 (0.1, 0.2, 0.3, and 0.4) units in $f_\lambda$~($V/I$).}
\label{fig:pg1015-result-04s020-spec}
\end{figure*}

In order to determine the misfit between the observations and the
model spectra for a given set of magnetic parameters, we
use the classical reduced \chisq\ as our penalty function. 
The optimisation problem of finding the best-fitting parameters is solved
using an evolutionary strategy, implemented by the \texttt{evoC} library 
(Trint \& Utecht 
1994)\footnote{ftp://biobio.bionik.tu-berlin.de/pub/software/evoC/}.
We assign equal statistical weight to the flux and polarization
spectra for all five rotational phases and to all wavelengths within
the individual spectra.
The statistical noise of the observations entering the \chisq\
function has been estimated by comparing the observed spectra after
the application of a Savitzky-Golay filter of \mbox{20\,\AA}\ width
with the unprocessed versions. 
The flux level of the model spectra has been adjusted to the
observations at the wavelengths marked with ticks at the top of
Fig.~\ref{fig:pg1015-obs-full} \textit{(right panel)}. This is an
attempt to remove differences between observed and model flux spectra
with wavelengths \mbox{$\ga 100$\,\AA}, as expected from the remaining
uncertainties in the detector response function. We do not correct
either for differences between the spectra on shorter wavelengths, nor
do we apply any correction to the polarization spectra. The resulting
\chisqred\ values are of the order of 80 indicating a gross
underestimation of the errors which enter the \chisq\ computation.
Larger observational errors and remaining systematic uncertainties in
the model spectra probably both contribute to this result. We have
experimented with different statistical weights assigned to different
wavelength regions -- as specific line features or subsections of the
continuum -- but found no convincing way to better define the goodness
of fit. We compromise on using the formal \chisqred, even if large,
as a guide line and decide by an admittedly subjective `by eye'
process which individual of similarly good fits to accept.

\begin{figure*}[t]
\includegraphics[bb=2 0 778 770,width=18cm,clip]{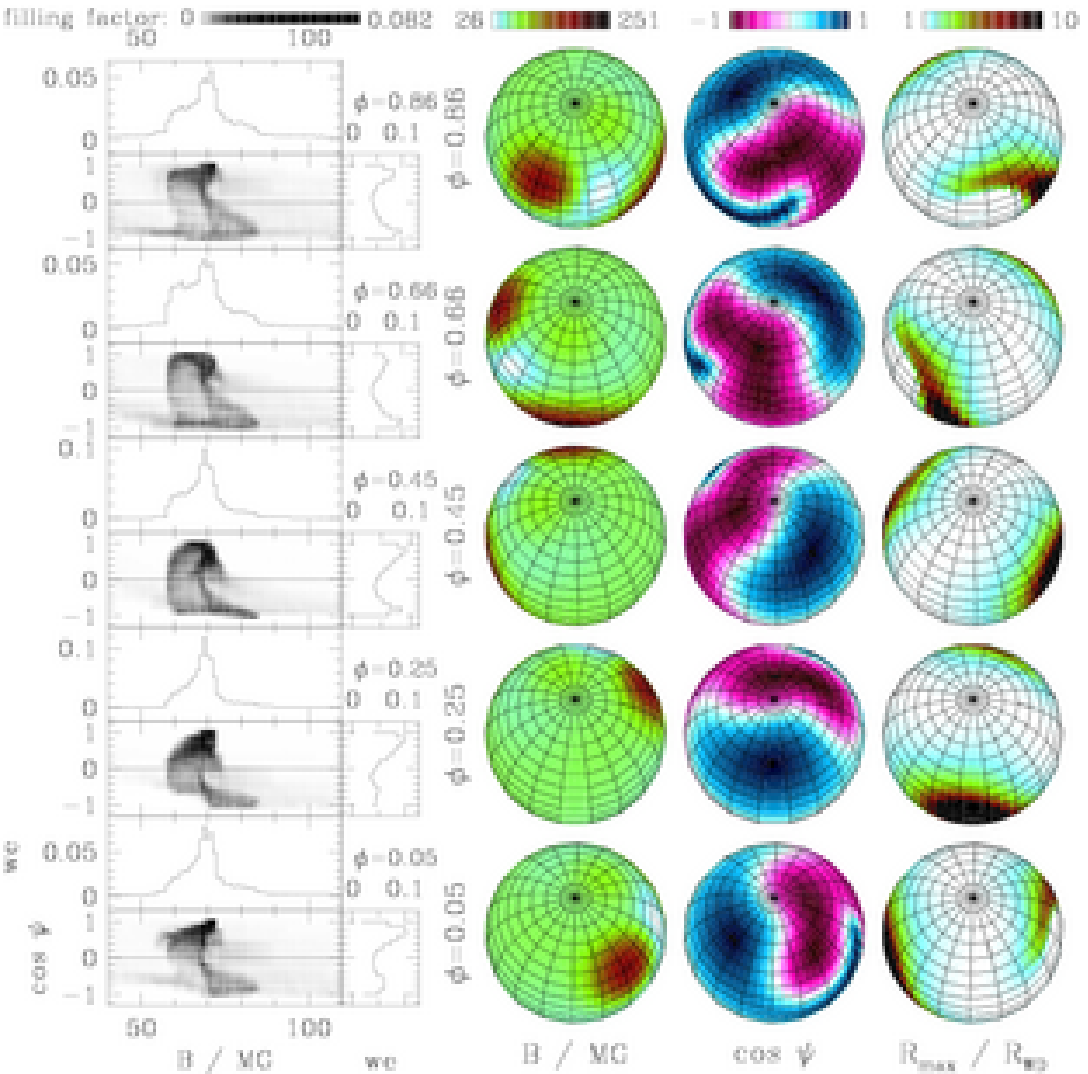} 
\caption{Zeeman tomographic analysis of the magnetic field
configuration of \opg\ using a superposition of three individually off-centred, non-aligned
dipoles.
\textit{Left panel:}
\bpd, \textit{right panel:} absolute value of the surface magnetic
field, cosine of the angle $\psi$ between the magnetic field direction
and the line of sight, and maximum radial distance reached by field
lines in units of the white dwarf radius.}
\label{fig:pg1015-result-04s020-bpsi}
\end{figure*}

\subsection{Field parametrization}
\label{sec:parametrization}

We have embarked on two different strategies in order to find an
adequate parametrization of the magnetic field geometry (see also
Paper I). The classical approach is that of an expansion of the scalar
magnetic potential into a series of spherical harmonics, characterised
by degree $l$ and order $m$ with \mbox{$m=0$} to $l$ for each $l$.  Given $l$
and $m$, two free parameters \glm\ and \hlm\ have to be specified, or
one parameter $g_l^0$ only for the zonal components with 
\mbox{$m = 0$} \citep{langel87-1}.  The approach is powerful, but is limited to
the description of rather simple structures if one truncates the
expansion at low values of $l$. Such a truncation is necessary in order
to avoid convergence problems of the fit, given the rapid growth in
the number of fit parameters which increases as $l(l+2)$ for the full
$l,m$-expansion. Conceptually simple structures, such as the sum of a
quadrupole and octupole with their axes inclined relative to each other,
cannot be realised if the series is truncated at low $l\ge3$. As an
alternative approach, we adopt a hybrid model consisting only of zonal
(\mbox{$m = 0$}) harmonics with independent tilt angles and off-centre
shifts. All tesseral components (\mbox{$m \neq 0$}) are ignored in
this case. Examples are, e.g., a combination of dipole, quadrupole and
octupole, and also a combination of three dipoles independently
inclined and offset from the centre.  Further details on the field
parametrizations are given in Papers~I and II.

In the following Section, we proceed systematically from simple
structures, as centred or off-centred single zonal components, over
the sum of such components, individually tilted and/or offset, to a
full multipole expansion truncated at $l=4$. The number of free
parameters varies between 4 and 27. In the case of the hybrid model,
the fit parameters for each component are the polar field strength
\mbox{$B_\mathrm{pol}$}, the tilt $\Theta$ and azimuth $\Phi$ of the
magnetic axis, and, if applicable, the offsets \xoff, \yoff, and
\zoff\ from the centre plus the inclination of the rotational axis
relative to the line of sight. For the truncated multipole model, we
fit the $l(l+2)$ parameters of the model, the two angles of the
reference axis, and the inclination.

\begin{figure*}[t]
\includegraphics[bb=32 168 564 690,width=18cm,clip]{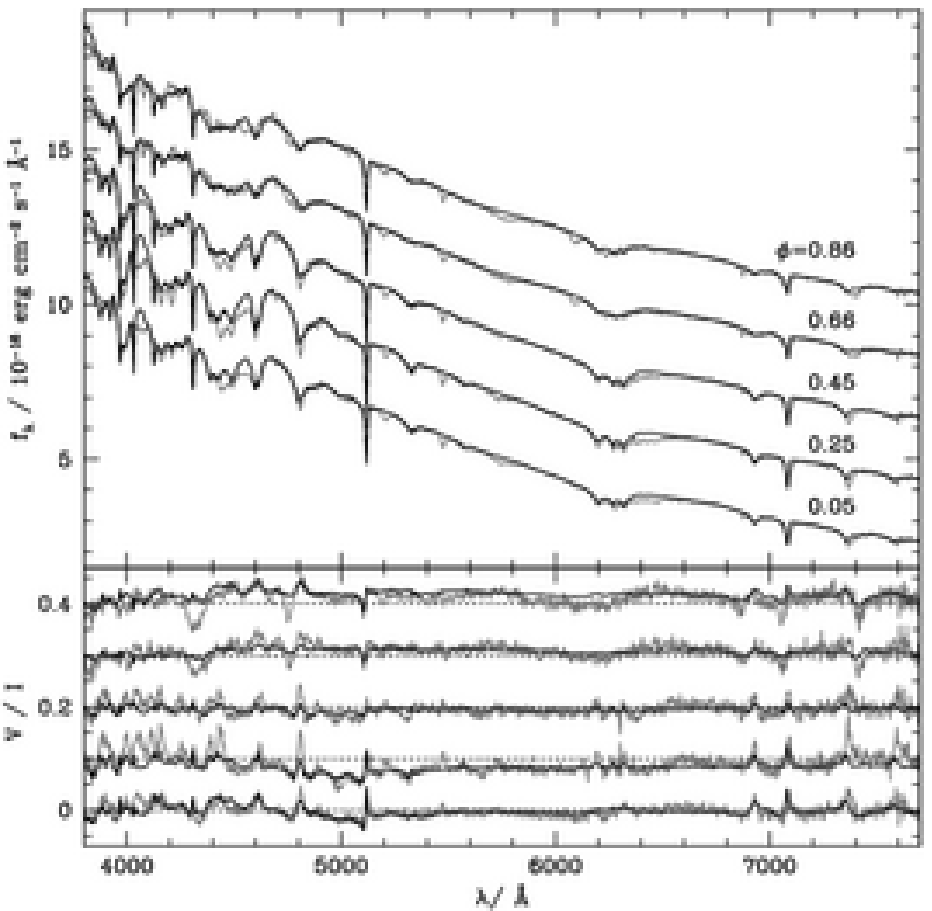} 
\caption{Observed spectra of \opg\ (grey curves) and best-fit synthetic spectra
(black curves) for a truncated multipole expansion up to
degree \mbox{$l = 4$}.
For clarity, the upper four flux (polarization) spectra have been shifted
upwards by 2, 4, 6, and 8 (0.1, 0.2, 0.3, and 0.4) units in
$f_\lambda$~($V/I$).}
\label{fig:pg1015-result-03s014a-spec}
\end{figure*}

\begin{figure*}[t]
\includegraphics[bb=2 0 778 770,width=18cm,clip]{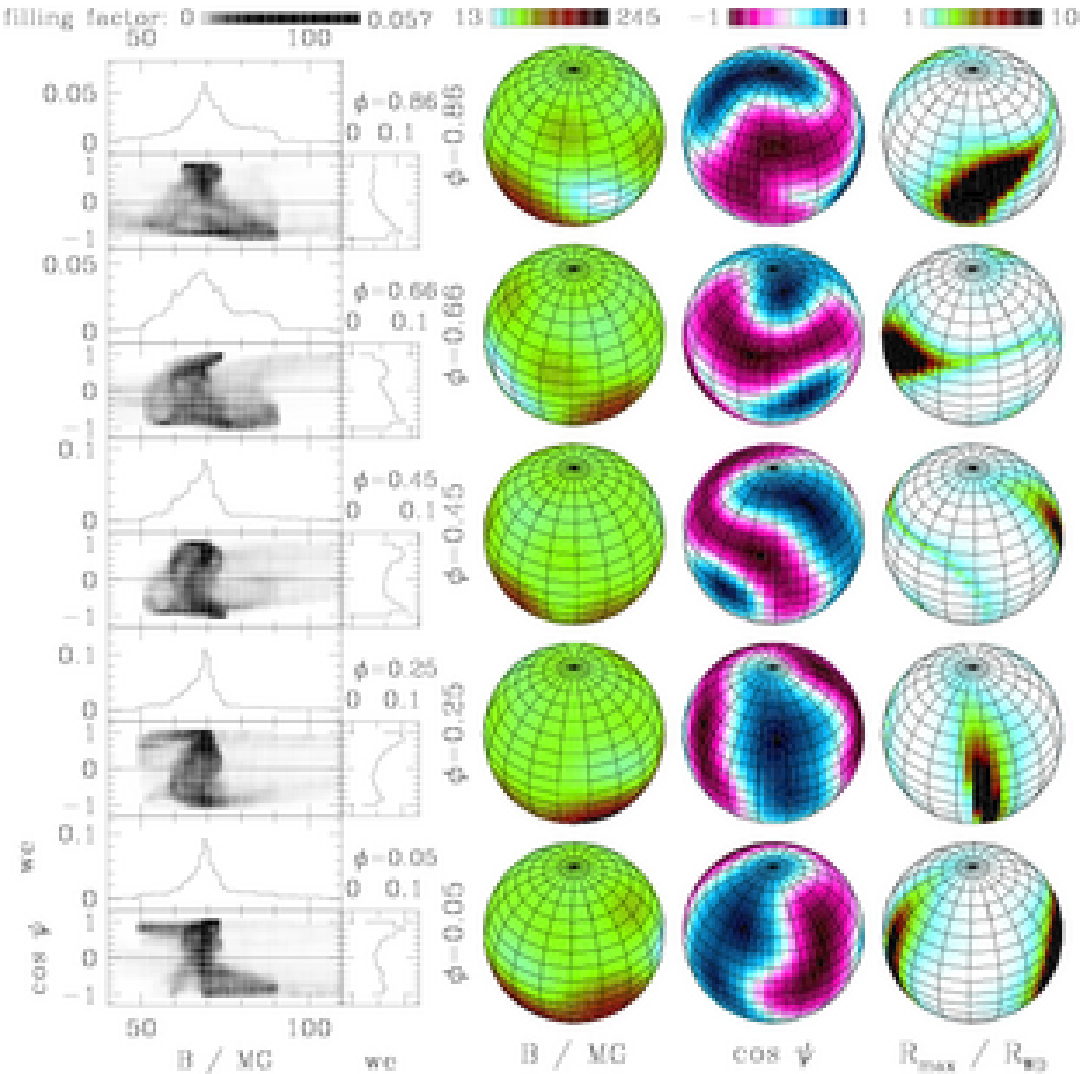} 
\caption{Zeeman tomographic analysis of the magnetic field
configuration of \opg\ using a truncated multipole expansion up to 
degree \mbox{$l = 4$}.
\textit{Left panel:}
\bpd, \textit{right panel:} absolute value of the surface magnetic
field, cosine of the angle $\psi$ between the magnetic field direction
and the line of sight, and maximum radial distance reached by field
lines in units of the white dwarf radius.}
\label{fig:pg1015-result-03s014a-bpsi}
\end{figure*}

\subsection{Results}
\label{sec:results}

For single centred dipoles, quadrupoles, or octupoles no satisfactory
fits could be obtained. Nevertheless, we note that the best fit with a
centred dipole to the single phase \mbox{$\phi = 0.25$} yielded a
value of \mbox{\bdip\ = 131\,$\pm$\,1\,MG} and an angle of
\mbox{$\alpha = 83$\degr}\ between the magnetic dipole axis and the
line of sight, which is compatible with the values quoted by
\citet{wickramasinghe+cropper88-1}. The model is far from optimal,
however, in particular for the other phases.

\begin{figure*}[t]
\includegraphics[width=18cm,clip]{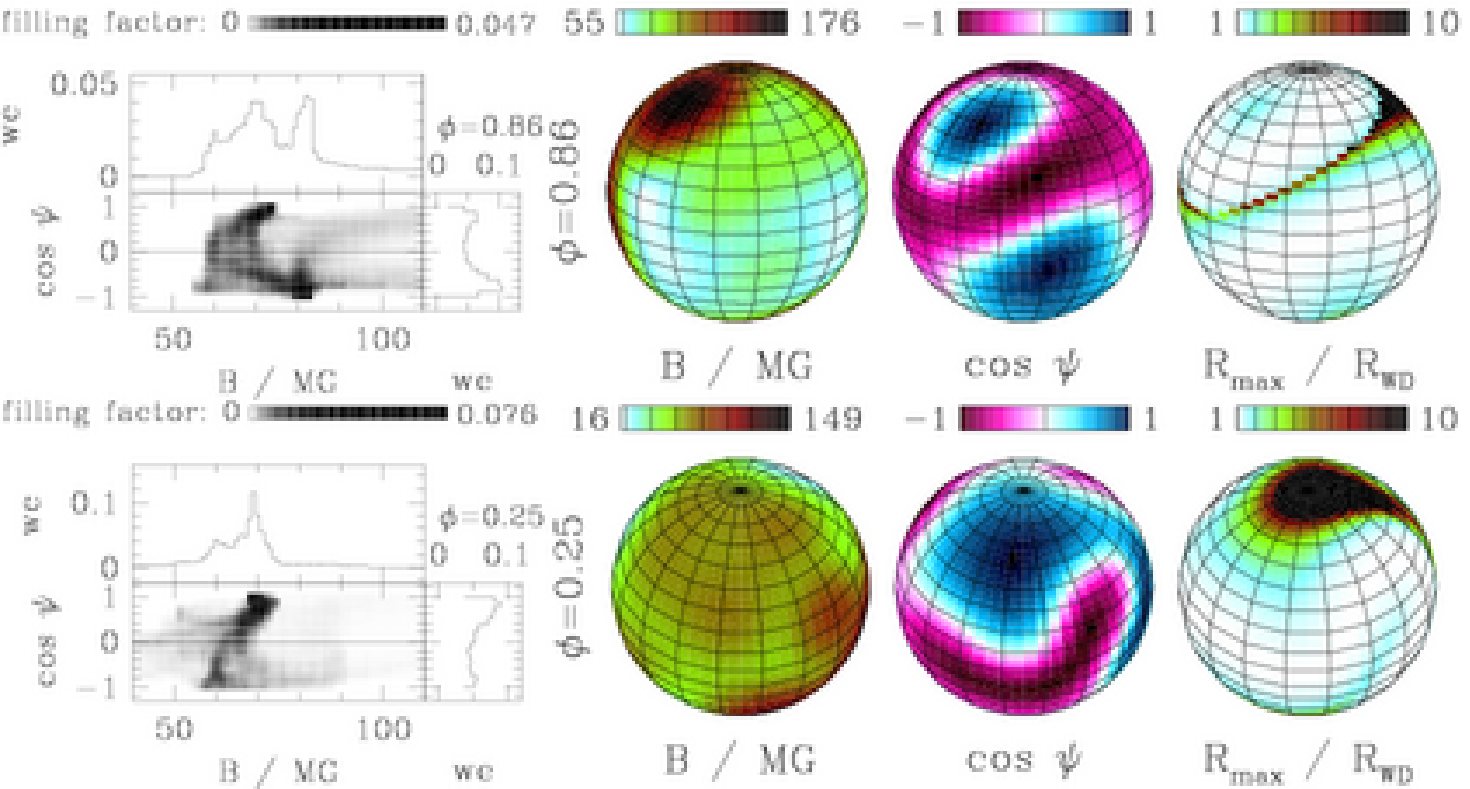}
\caption{Best-fit results for individual single phase fits to the 
Zeeman maximum (\mbox{$\phi = 0.25$}, \textit{bottom panels}) and minimum
(\mbox{$\phi = 0.86$}, \textit{top panels}) phases. The field has been 
parametrized 
by a truncated multipole expansion up to degree \mbox{$l = 5$} for 
\mbox{$\phi = 0.25$},
and up to \mbox{$l = 4$} for \mbox{$\phi = 0.86$}.
\textit{Left panels:}
\bpd, \textit{right panels:} absolute value of the surface magnetic
field, cosine of the angle $\psi$ between the magnetic field direction
and the line of sight, and maximum radial distance reached by field
lines in units of the white dwarf radius.}
\label{fig:pg1015-result-singlephase}
\end{figure*}

In Fig.~\ref{fig:pg1015-multisp} we show the observed and best-fitting
model circular polarizations at \mbox{$\phi = 0.25$} and 0.86 for
selected hybrid and multipole parametrizations with increasing
complexity.  The simplest configuration is that of an offset dipole
(denoted by D~offs in Fig.~\ref{fig:pg1015-multisp}).  The
corresponding values of the magnetic parameters are listed in
Table~\ref{tab:pg1015-fit-parameters}. The
overall shape of the continuum polarization is reproduced fairly well
for all phases, but the model fails seriously in the spectral lines,
especially for \mbox{$\phi = 0.86$}.
For an offset quadrupole, the result was comparably poor. 
An offset octupole fared slightly better, but still was
considered unsatisfactory.
While acceptable fits could be obtained with offset dipoles and
quadrupoles for \ohe\  (Paper~II), the magnetic field geometry of \opg\
seems to be significantly more complex.

As a next step, 
we proceed to the same hybrid field models that yielded successful
fits for \ohe. These are produced by the superposition of dipole, quadrupole, 
and octupole components and the introduction of a \emph{common} off-centre shift.
For the nonaligned centred dipole--quadrupole--octupole combination
(DQO~na~ctrd in Fig.~\ref{fig:pg1015-multisp}), the polarization
spectrum at \mbox{$\phi = 0.86$} is reproduced remarkably well, in
particular the broad negative dips in polarization at 4310 and
\mbox{4740\,\AA}, and the \mbox{2s\tz\,$\rightarrow$\,3p\tn}\
transition at \mbox{7085\,\AA}.
The frequency distribution of magnetic field strengths and directions
(\bpd, not shown) is double-peaked at fields of 69 and \mbox{81\,MG} 
in accordance with the considerations of
Sect.~\ref{sec:qualitative}. 
The essence, however, is that the
reproduction is poor for the other phases, including 
\mbox{$\phi = 0.25$}, although \mbox{$\phi = 0.86$} fits well.
Introducing a common offset to the model (DQO~na~offs in
Fig.~\ref{fig:pg1015-multisp}) improves the fit at \mbox{$\phi =
0.25$}, but shows significant deviations at \mbox{$\phi = 0.86$}.
As can be seen from the best-fit magnetic parameters in lines~(2) and (3)
of Table~\ref{tab:pg1015-fit-parameters}, both the centred and the
shifted hybrid model are dominated by the octupole component, again
suggesting a rather complex geometry.
We conclude that the models considered so far still do not suffice to
achieve a fair reproduction at all phases.

In an attempt to improve the fits, we stayed with hybrid models and
tried to model a star with several field concentrations by adding
three dipoles which are \emph{individually} tilted and
off-centred. The model fares surprisingly well despite the absence of
quadrupole and octupole components. We conclude that the ability to
place and adjust three spots meets reality rather closely. The fit
results for this `triple dipole' model are displayed in
Fig.~\ref{fig:pg1015-result-04s020-spec},
Fig.~\ref{fig:pg1015-result-04s020-bpsi}, and
Table~\ref{tab:pg1015-fit-parameters-ddd} and are discussed below.

Finally, we considered a truncated multipole expansion up to \mbox{$l
= 4$} which includes all orders with indices \mbox{$m=0$} to $l$. This
`truncated multipole' model achieves a \chisqred\ value similarly good
as the triple dipole model and both provide substantially better
fits than the parametrizations discussed above if all rotational
phases are considered. In particular, the wavelength-dependent level of the
continuum circular polarization 
is reproduced more accurately.  The results for the truncated
multipole model are shown in
Figs.~\ref{fig:pg1015-result-03s014a-spec} and
\ref{fig:pg1015-result-03s014a-bpsi} and in
Table~\ref{tab:pg1015-fit-parameters-multi}.

Although both magnetic field topologies are generated by entirely different 
parametrizations 
(Tables~\ref{tab:pg1015-fit-parameters-ddd} and 
\ref{tab:pg1015-fit-parameters-multi}), 
the overall appearance of their \bpds\ is similar.
For both configurations, the visible part of the surface
field is dominated by an extended area with a rather
small variation of the field strength that leads to the pronounced
peaks at \mbox{$\simeq$\,70\,MG} in the \bpds.
The phases \mbox{$\phi = 0.25$} and 0.45 are almost entirely dominated
by this region of constant (although not homogenous) field. For the
remaining phases, the low- and high-field tails of the field strength
distribution become more prominent.
In the triple dipole model, the regions with low and high fields
become manifest as small spots, one of which reaches up to
\mbox{$\simeq$\,160\,MG}.
A second high-field
region appears at the stellar limb for phases
\mbox{$\phi = 0.66$} and 0.86.
For the truncated multipole model, a similarly prominent 
high-field spot is not seen, but a low-field spot appears close
to the stellar limb for \mbox{$\phi = 0.66$} and 0.86 and finds its
counterpart in the hybrid model.
The high-field regions are divided into two small spots with field
values up to \mbox{$\simeq$\,90\,MG} and pronounced negative
\mbox{$\cos \psi$}, leading to a distinctive signature in the \bpd,
and two narrow areas at the stellar limb with $B$ up to
\mbox{$\simeq$\,150\,MG} and \mbox{245\,MG}, which belong to
high-field spots on the hidden part of the stellar surface. This
comparison of the triple dipole model and the truncated multipole
model emphasises the fact that the principal information on the field
structure is contained in the \bpd s. These diagrams indicate that the
distributions of the magnetic field are similar, a conclusion which is
impossible to draw from the numerical descriptions in
Tables~\ref{tab:pg1015-fit-parameters-ddd} and \ref
{tab:pg1015-fit-parameters-multi}.

For both configurations substantial deviations from the
observational data remain, e.g., at the broad dips in $V/I$ at 
4310 and \mbox{4740\,\AA}\ for phases \mbox{$\phi = 0.66$} and 0.86 which
both models cannot reproduce correctly.
While it is quite likely that a larger number of free parameters --
and, hence, a still further increased complexity of the field -- will
improve the fits, our present optimisation procedure cannot handle
more free parameters and prevents us to pursue this possibility
further. As a consequence, it remains unclear what part of the
remaining discrepancies, if any, may be due to systematic errors in
the model spectra.

A possible way to answer the last question at least in part is to
perform fits to the flux and polarization spectra for a 
\emph{single} phase and abandon the requirement that
the model should simultaneously fit the other phases. The resulting
model may be wrong in a global sense, but will provide a more accurate
description of the field distribution over the visible hemisphere at
the selected phase.  Figure~\ref{fig:pg1015-result-singlephase} shows
the resulting \bpds\ at \mbox{$\phi = 0.25$} and \mbox{$\phi = 0.86$}
using the truncated multipole model with $l=5$ and $l=4$,
respectively. 
The corresponding polarization spectra are shown at the bottom of
Fig.~\ref{fig:pg1015-multisp}.
The \bpds\ differ, in fact, from the
distributions for the same phases obtained from the simultaneous fits
to all five phases in Figs.~\ref{fig:pg1015-result-04s020-bpsi} and
\ref{fig:pg1015-result-03s014a-bpsi}. The most obvious new feature is
the appearance of a second field maximum at 82\,MG for \mbox{$\phi =
0.86$}. At \mbox{$\phi = 0.25$}, the distributions of the field
strengths look similar, but the $\psi$-distributions and, hence, the
field geometries differ (see
Figs.~\ref{fig:pg1015-result-04s020-bpsi},
\ref{fig:pg1015-result-03s014a-bpsi}, and
\ref{fig:pg1015-result-singlephase}). We conclude that a major
fraction of the misfits still present in our two best global fits is
due to the disability of the models to appropriately account for the
complexity of the field. The parameter-free spectral synthesis of
\citet{donatietal94-1} would allow to optimise the fit to a single
phase still further and answer the quest for the best possible fit
with the present database, at the expense of a global field solution,
however (see also Section~5).

We conclude that the field models used by us are barely sufficient to
describe a single phase of the observations of \opg, and are certainly
not complex enough to describe the global field configuration by
fits to all five phases.

Several potential sources of deviations
between observations and synthetic model spectra have been proposed in 
Paper~II and arise both from remaining uncertainties on the observational 
(flat-fielding, flux calibration)
and on the theoretical side.
Since \opg\ is dominated by substantially higher magnetic field
strengths than \ohe, additional error sources for the theoretical
model spectra have to be considered.
With growing magnetic field strength, for instance, 
the influence of electric fields
on line positions and strengths becomes increasingly important.
For the case of arbitrarily oriented electric and magnetic fields
no discrete symmetry is left, 
leading to slightly different transition wavelengths and oscillator
strengths than those computed for the diamagnetic case,
and even to the occurrence of additional dipole transitions
\citep{fassbinder+schweizer96-1,burleighetal99-1}. 
Lines of such type have not been included in the atomic data tables
used for the computation of our synthetic model spectra and,
consequently, cannot be reproduced by our fits. They are possibly
responsible for the sharp line features at \mbox{5200\,\AA}\ and 
\mbox{5475\,\AA}\ and
the washed out feature at \mbox{5750\,\AA}, which are not explained by our
models (Figs.~\ref{fig:pg1015-result-04s020-spec} and
\ref{fig:pg1015-result-03s014a-spec}).

Another theoretical uncertainty arises from the simplified treatment of
the field-dependent bound-free and free-free transitions as described in 
\citet{jordan92-1}.
For the case of \ogrw\ with \mbox{$B \simeq 320$\,MG},
\citet{jordan+merani95-1} have shown that a more consistent but
numerically extremely expensive treatment of the continuum opacities
can yield slightly different results for the polarization.
We regard the uncertainties arising from both potential error sources
as relatively small at fields of \mbox{$\simeq$\,80\,MG}, but slight
deviations from our synthetic spectra cannot be ruled out.


\section{Discussion}
\label{sec:discussion}

In this work, we have analysed high-quality spectropolarimetric
data of \opg\ covering a whole rotational period with the Zeeman tomography
code described in Papers~I and II.
We have achieved good fits to the observations, but require a magnetic field
geometry that is significantly more complex than the 
popular assumption of centred or moderately offset dipoles and quadrupoles
proposed by several authors in the past
\citep[][ and references therein]{wickramasinghe+ferrario00-1}.
In fact, the magnetic field structure of \opg\ is even more complex than that
derived for \ohe\ (Paper~II)
and can be successfully modelled by two different parametrizations:
On the one hand, we used a 
hybrid model of off-centred, tilted zonal (\mbox{$m = 0$}) components, 
and on the other hand,
a truncated multipole expansion including all \mbox{$m \ne 0$} components.
Hence, our results further confirm the evidence that the magnetic field 
structures of MWDs are non-trivial and require higher multipole components for 
an accurate description. We have shown, in particular, that a simple oblique
dipole model as devised by 
\citet{wickramasinghe+cropper88-1} does not suffice
to describe the complex
Zeeman absorption features.

Our findings for \opg\ show the difficulties that are inherent to the
description of a star's complex magnetic field geometry with only a few
numeric parameters.
For this object, an appropriate description requires to quote at least
the range of field strengths that contributes effectively to the
spectral shape, 50 to \mbox{90\,MG}, and the phase-dependent maxima of
the field strength distributions at 69 and \mbox{82\,MG}.

For both parametrization strategies the number of free parameters
required for a good fit approaches the maximum number our tomography
algorithm can currently handle, and we could reach equivalent quality
levels for the best fits.
The fits are sufficiently good to be certain that the best-fitting
field geometry comes reasonably close to reality, but we could not
reach the same high quality of the fits as for \ohe\ (Paper~II).

We propose two explanations for the remaining differences between the
observations and our best-fitting models:
(i) systematic uncertainties in the model spectra arising already in
the field regime at \mbox{$B \ga 50$\,MG}, as discussed at the end of
Sect.~\ref{sec:results};
(ii) insufficient spatial resolution of the magnetic field
distribution provided by our field geometry models.
Given the limited number of free field parameters and the corresponding
limitations regarding the attainable level of complexity in the \bpds,
residuals caused by both effects cannot be disentangled.
However, the fact that fits for a single phase were clearly superior to the 
simultaneous fits
for all five phases suggests that there is room for improvement
on the side of finer-grained field distributions.
Hence, it would be necessary to examine the quality of fits to all phases
with an increased number of free parameters before a reliable estimation of
the effects of 
systematic errors in the synthetic model spectra
becomes possible.

A different method to tackle this problem would be an approach
like that of \citet{donatietal94-1}, who
optimised directly the frequency distribution of field strengths and
directions. Such an approach has the advantage that a formal
best fit to the observed spectra is found which 
shows the smallest residuals of all possible
combinations of database spectra, but a few potential traps should be kept
in mind: (i) there is no unique relation between the \bpd\ and the field
topology (see Paper~I for an example of ambiguous field 
configurations);
(ii) there is no guarantee
that the derived \bpd\ corresponds to a field which fits in a
globally consistent picture if all rotational phases are regarded;
(iii) and it cannot be guaranteed 
that a distribution of electric currents 
exists which produces the derived \bpd. 
Remaining discrepancies between the observations and 
integrated synthetic spectra
derived with the method of \citet{donatietal94-1} would 
be, nevertheless, a good measure for the magnitude of systematic errors in the 
model spectra.

Somewhat unfortunately, the two objects we have analysed so far with
our Zeeman tomography code both have equal \teff\ (Paper II and this
work). It would be desirable to examine the magnetic field geometries
of MWDs for a broader range of effective temperatures in order to
search for potential effects of a secular field evolution as a
function of cooling age.

The outcome of the first three data releases of the \sdss\ (SDSS) has nearly
tripled the number of known MWDs.
The multitude of newly found objects covers a broad range of effective
temperatures and surface dipole field strengths
\citep{gaensickeetal02-1,schmidtetal03-1,vanlandinghametal05-1}. Several
new objects with \mbox{$B>200$\,MG} have been found, while the
majority of objects is found in the low-field regime with
\mbox{$B<20$\,MG}. The SDSS objects provide a vast hunting ground for
further systematic studies of the field
geometry of MWDs.
The SDSS has also nearly doubled the number of helium-rich
MWDs and an interesting option is the extension of our Zeeman
tomography technique to the synthesis of their spectra.
Calculations of the atomic parameters for He are available now and
first applications to helium-rich MWDs are available
\citep{jordanetal01-1,wickramasingheetal02-1}.

Another promising avenue of research is the study of accreting MWDs in close
mass-transferring binaries. In the following papers of this series, we will
investigate the magnetic field structures of the accreting MWDs in
magnetic cataclysmic variables.


\begin{acknowledgements}
This work was supported in part by BMBF/DLR grant
\mbox{50\,OR\,9903\,6}.  BTG was supported by a PPARC Advanced
Fellowship.
\end{acknowledgements}


\bibliographystyle{aa}

\end{document}